\documentstyle[12pt]{art12}
\makeatletter
\let\reset@font\empty

\def\indexname{Index}
\def\figurename{Figure}
\def\tablename{Table}
\def\abstractname{Abstract}

\def\@ptsize{0}
\@namedef{ds@11pt}{\def\@ptsize{1}}
\@namedef{ds@12pt}{\def\@ptsize{2}}
\def\ds@twoside{\@twosidetrue
           \@mparswitchtrue}

\def\ds@draft{\overfullrule 5\p@}

\newif\if@titlepage \@titlepagefalse
\def\ds@titlepage{\@titlepagetrue}

\def\ds@twocolumn{\@twocolumntrue}

\newdimen\mathindent
\newif\ifletter
\newif\ifpmb
\newlength{\varind}
\newlength{\figdepth}
\newlength{\figwidth}
\newlength{\secfigwidth}

\newlength{\indentedwidth}

\newcounter{jnl}
\newcounter{yr}
\newcounter{tabtype}
\newcounter{figtype}
\newcounter{eqnval}
\@twosidetrue
\def\ds@draft{\overfullrule 5\p@}
\@options

\def\@normalsize{\@setsize\normalsize{16pt}\xiipt\@xiipt
  \abovedisplayskip 12pt plus3pt minus6pt
  \belowdisplayskip \abovedisplayskip
  \abovedisplayshortskip \z@ plus4pt
  \belowdisplayshortskip 7pt plus4pt minus4pt}
\def\small{\@setsize\small{14pt}\xipt\@xipt
  \abovedisplayskip 10pt plus 3pt minus 4pt
  \belowdisplayskip \abovedisplayskip
  \abovedisplayshortskip \z@ plus3pt
  \belowdisplayshortskip 5pt plus3pt minus 3pt
  \def\@listi{\topsep 5pt plus 3pt minus 3pt\parsep 0pt plus 1pt
         \itemsep \parsep}}
\def\footnotesize{\@setsize\footnotesize{14pt}\xpt\@xpt
  \abovedisplayskip 7pt plus 3pt minus 4pt
  \belowdisplayskip \abovedisplayskip
  \abovedisplayshortskip \z@ plus 2pt
  \belowdisplayshortskip 3pt plus 1pt minus2pt
  \def\@listi{\topsep 4pt plus 2pt minus 2pt\parsep 0pt plus 1pt
         \itemsep \parsep}}
\def\scriptsize{\@setsize\scriptsize{13pt}\ixpt\@ixpt}
\def\tiny{\@setsize\tiny{10pt}\viipt\@viipt}
\def\large{\@setsize\large{18pt}\xivpt\@xivpt}
\def\Large{\@setsize\Large{22pt}\xviipt\@xviipt}
\def\LARGE{\@setsize\LARGE{25pt}\xxpt\@xxpt}
\def\huge{\@setsize\huge{30pt}\xxvpt\@xxvpt}
\def\Huge{\@setsize\Huge{30pt}\xxvpt\@xxvpt}
\normalsize
\skip\footins 12\p@ plus 4\p@ minus 2\p@
\@beginparpenalty -\@lowpenalty
\@endparpenalty -\@lowpenalty
\@itempenalty -\@lowpenalty

\@noskipsecfalse   

\def\section{\@startsection{section}{1}{\z@}{-3.5ex plus -1ex minus
 -.2ex}{2.3ex plus .2ex}{\noindent\reset@font\normalsize\bf\raggedright}}
\def\subsection{\@startsection{subsection}{2}{\z@}{-3.25ex plus -1ex minus
 -.2ex}{1.5ex plus .2ex}{\noindent\reset@font
  \normalsize\it\raggedright\nohyphens}}
\def\subsubsection{\@startsection{subsubsection}{3}{\z@}{-3.25ex plus
-1ex minus -.2ex}{-1em}{\reset@font\normalsize\it\nohyphens}}
\def\paragraph{\@startsection
 {paragraph}{4}{\z@}{3.25ex plus 1ex minus
.2ex}{-1em}{\reset@font\normalsize\it}}
\def\subparagraph{\@startsection
 {subparagraph}{4}{\parindent}{3.25ex plus 1ex minus
 .2ex}{-1em}{\reset@font\normalsize\it}}

\def\@sect#1#2#3#4#5#6[#7]#8{\ifnum #2>\c@secnumdepth
     \let\@svsec\@empty\else
     \refstepcounter{#1}\edef\@svsec{\csname the#1\endcsname.\hskip 1em}\fi
     \@tempskipa #5\relax
      \ifdim \@tempskipa>\z@
        \begingroup #6\relax
          \noindent{\hskip #3\relax\@svsec}{\interlinepenalty \@M #8\par}%
        \endgroup
       \csname #1mark\endcsname{#7}\addcontentsline
         {toc}{#1}{\ifnum #2>\c@secnumdepth \else
                      \protect\numberline{\csname the#1\endcsname}\fi
                    #7}\else
        \def\@svsechd{#6\hskip #3\relax  
                   \@svsec #8\csname #1mark\endcsname
                      {#7}\addcontentsline
                           {toc}{#1}{\ifnum #2>\c@secnumdepth \else
                             \protect\numberline{\csname the#1\endcsname}\fi
                       #7}}\fi
     \@xsect{#5}}
\def\@ssect#1#2#3#4#5{\@tempskipa #3\relax
   \ifdim \@tempskipa>\z@
     \begingroup #4\noindent{\hskip #1}{\interlinepenalty
   \@M #5\par}\endgroup
   \else \def\@svsechd{#4\hskip #1\relax #5}\fi
    \@xsect{#3}}

\def\appendix{\@@par
 \setcounter{section}{0}
 \setcounter{subsection}{0}
 \setcounter{subsubsection}{0}
 \setcounter{equation}{0}
 \setcounter{figure}{0}
 \setcounter{table}{0}
 \def\thesection{Appendix \Alph{section}}
 \def\theequation{\ifnumbysec
      \Alph{section}.\arabic{equation}\else
      \Alph{section}\arabic{equation}\fi}
 \def\thetable{\ifnumbysec
      \Alph{section}\arabic{table}\else
      A\arabic{table}\fi}
 \def\thefigure{\ifnumbysec
      \Alph{section}\arabic{figure}\else
      A\arabic{figure}\fi}}

\leftmargin\leftmargini
\labelwidth\leftmargini\advance\labelwidth-\labelsep
\def\@listI{\leftmargin\leftmargini \parsep 4\p@ plus2\p@ minus\p@
\topsep 8\p@ plus2\p@ minus4\p@
\itemsep 4\p@ plus2\p@ minus\p@}

\let\@listi\@listI
\@listi

\def\@listii{\leftmargin\leftmarginii
 \labelwidth\leftmarginii\advance\labelwidth-\labelsep
 \topsep 3\p@ plus 1\p@ minus 1\p@
 \parsep 0\p@ plus 1\p@
 \itemsep \parsep}
\def\@listiii{\leftmargin\leftmarginiii
 \labelwidth\leftmarginiii\advance\labelwidth-\labelsep
 \topsep 2\p@ plus 1\p@ minus 1\p@
 \parsep \z@ \partopsep 1\p@ plus 0\p@ minus 1\p@
 \itemsep \topsep}
\def\@listiv{\leftmargin\leftmarginiv
 \labelwidth\leftmarginiv\advance\labelwidth-\labelsep}
\def\@listv{\leftmargin\leftmarginv
 \labelwidth\leftmarginv\advance\labelwidth-\labelsep}
\def\@listvi{\leftmargin\leftmarginvi
 \labelwidth\leftmarginvi\advance\labelwidth-\labelsep}

\def\hexnumber@#1{\ifcase#1 0\or 1\or 2\or 3\or 4\or 5\or 6\or 7\or 8\or
 9\or A\or B\or C\or D\or E\or F\fi}
\edef\bffam@{\hexnumber@\bffam}
\mathchardef\bGamma "0\bffam@00
\mathchardef\bDelta "0\bffam@01
\mathchardef\bTheta "0\bffam@02
\mathchardef\bLambda "0\bffam@03
\mathchardef\bXi "0\bffam@04
\mathchardef\bPi "0\bffam@05
\mathchardef\bSigma "0\bffam@06
\mathchardef\bUpsilon "0\bffam@07
\mathchardef\bPhi "0\bffam@08
\mathchardef\bPsi "0\bffam@09
\mathchardef\bOmega "0\bffam@0A

\def\theenumi{\roman{enumi}}

\def\theenumii{\alph{enumii}}
\def\p@enumii{\theenumi.}

\def\theenumiii{\arabic{enumiii}}
\def\p@enumiii{\p@enumii.\theenumii}

\def\p@enumiv{\p@enumiii.\theenumiii}

\def\labelitemi{$\m@th\bullet$}

\def\labelitemiii{$\m@th\ast$}
\def\labelitemiv{$\m@th\cdot$}

\def\verse{\let\\=\@centercr
 \list{}{\itemsep\z@ \itemindent -1.5em\listparindent \itemindent
 \rightmargin\leftmargin\advance\leftmargin 1.5em}\item[]}

\def\quotation{\list{}{\listparindent 1.5em
 \itemindent\listparindent
 \rightmargin\leftmargin\parsep 0\p@ plus 1\p@}\item[]}

\def\descriptionlabel#1{\hspace\labelsep \bf #1}
\def\description{\list{}{\labelwidth\z@ \itemindent-\leftmargin
 \let\makelabel\descriptionlabel}}

\def\enumerate{\ifnum \@enumdepth >3 \@toodeep\else
      \advance\@enumdepth \@ne
      \edef\@enumctr{enum\romannumeral\the\@enumdepth}\list
      {\csname label\@enumctr\endcsname}{\usecounter
        {\@enumctr}\def\makelabel##1{##1\hss}}\fi}
\def\itemize{\ifnum \@itemdepth >3 \@toodeep\else \advance\@itemdepth \@ne
\edef\@itemitem{labelitem\romannumeral\the\@itemdepth}%
\list{\csname\@itemitem\endcsname}{\def\makelabel##1{##1\hss}\topsep=3pt
  \parsep=0pt\listparindent=0pt\itemsep=0pt\partopsep=0pt\rightmargin=0pt
  }\fi}
\tabbingsep \labelsep
\skip\@mpfootins = \skip\footins
\def\titlepage{\@restonecolfalse\if@twocolumn\@restonecoltrue\onecolumn
     \else \newpage \fi \thispagestyle{myheadings}\c@page\z@}

\def\endtitlepage{\if@restonecol\twocolumn \else \newpage \fi}

\newcounter {section}
\newcounter {subsection}[section]
\newcounter {subsubsection}[subsection]
\newcounter {paragraph}[subsubsection]
\newcounter {subparagraph}[paragraph]

\def\thesection {\arabic{section}}

\def\@chapapp{Section}

\def\@pnumwidth{1.55em}
\def\@tocrmarg {2.55em}
\def\@dotsep{4.5}
\def\tableofcontents{\@restonecolfalse\if@twocolumn\@restonecoltrue
 \onecolumn\fi\section*{Contents}{}\thispagestyle{empty}
 \@starttoc{toc}\if@restonecol\twocolumn\fi}
\def\l@section{\@dottedtocline{1}{1.5em}{2.3em}}
\def\l@subsection{\@dottedtocline{2}{3.8em}{3.2em}}
\def\l@subsubsection{\@dottedtocline{3}{7.0em}{4.1em}}
\def\l@paragraph{\@dottedtocline{4}{10em}{5em}}
\def\l@subparagraph{\@dottedtocline{5}{12em}{6em}}
\def\listoffigures{\@restonecolfalse\if@twocolumn\@restonecoltrue\onecolumn
 \fi\section*{List of Figures\@mkboth
 {LIST OF FIGURES}{LIST OF FIGURES}}\@starttoc{lof}\if@restonecol\twocolumn
 \fi}
\def\l@figure{\@dottedtocline{1}{1.5em}{2.3em}}
\def\listoftables{\@restonecolfalse\if@twocolumn\@restonecoltrue\onecolumn
 \fi\section*{List of Tables\@mkboth
 {LIST OF TABLES}{LIST OF TABLES}}\@starttoc{lot}\if@restonecol\twocolumn
 \fi}
\let\l@table\l@figure
\def\@dottedtocline#1#2#3#4#5{\ifnum #1>\c@tocdepth \else
  \vskip \z@ plus .2\p@
  {\leftskip #2\relax \rightskip \@tocrmarg \parfillskip -\rightskip
    \parindent #2\relax\@afterindenttrue
   \interlinepenalty\@M
   \leavevmode
   \@tempdima #3\relax \advance\leftskip \@tempdima
   \hbox{}\hskip -\leftskip
    #4\nobreak\hfill \nobreak \hbox to\@pnumwidth{\hfil
   \rm #5}\@@par}\fi}

\@addtoreset{footnote}{page}
\long\def\@makefntext#1{\parindent 1em\noindent
 \makebox[1em][l]{\footnotesize\rm$\m@th{\fnsymbol{footnote}}$}%
 \footnotesize\rm #1}
\def\@makefnmark{\hbox{${\fnsymbol{footnote}}\m@th$}}
\def\@thefnmark{\fnsymbol{footnote}}
\def\footnote{\@ifnextchar[{\@xfootnote}{\stepcounter{\@mpfn}%
       \begingroup\let\protect\noexpand
       \xdef\@thefnmark{\thempfn}\endgroup
     \@footnotemark\@footnotetext}}
\def\@fnsymbol#1{\ifcase#1\or \dagger\or \ddagger\or \S\or
   \|\or \P\or ^{+}\or ^{\tsty *}\or \sharp
   \or \dagger\dagger \else\@ctrerr\fi\relax}

\def\[{\relax\ifmmode\@badmath\else
 \begin{trivlist}
 \@beginparpenalty\predisplaypenalty
 \@endparpenalty\postdisplaypenalty
 \item[]\leavevmode
 \hbox to\linewidth\bgroup$ \displaystyle
 \hskip\mathindent\bgroup\fi}
\def\]{\relax\ifmmode \egroup $\hfil \egroup \end{trivlist}\else \@badmath \fi}
\def\equation{\@beginparpenalty\predisplaypenalty
 \@endparpenalty\postdisplaypenalty
\refstepcounter{equation}\trivlist \item[]\leavevmode
 \hbox to\linewidth\bgroup $ \displaystyle
\hskip\mathindent}
\def\endequation{$\hfil \displaywidth\linewidth\@eqnnum\egroup \endtrivlist}
\def\eqnarray{\stepcounter{equation}\let\@currentlabel=\theequation
\global\@eqnswtrue
\global\@eqcnt\z@\tabskip\mathindent\let\\=\@eqncr
\abovedisplayskip\topsep\ifvmode\advance\abovedisplayskip\partopsep\fi
\belowdisplayskip\abovedisplayskip
\belowdisplayshortskip\abovedisplayskip
\abovedisplayshortskip\abovedisplayskip
$$\halign to
\linewidth\bgroup\@eqnsel$\displaystyle\tabskip\z@
 {##{}}$&\global\@eqcnt\@ne $\displaystyle{{}##{}}$\hfil    
 &\global\@eqcnt\tw@ $\displaystyle{{}##}$\hfil
 \tabskip\@centering&\llap{##}\tabskip\z@\cr}
\def\endeqnarray{\@@eqncr\egroup
 \global\advance\c@equation\m@ne$$\global\@ignoretrue }
\newcommand{\jl}[1]{\setcounter{jnl}{#1}%
    \ifnum\thejnl=12\global\pmbtrue\fi
    \ifnum\thejnl=15\global\pmbtrue\fi}
\def\journal{\ifnum\thejnl=1 J. Phys.\ A: Math.\ Gen.\
        \else\ifnum\thejnl=2 J. Phys.\ B: At.\ Mol.\ Opt.\ Phys.\
        \else\ifnum\thejnl=3 J. Phys.:\ Condens. Matter\
        \else\ifnum\thejnl=4 J. Phys.\ G: Nucl.\ Part.\ Phys.\
        \else\ifnum\thejnl=5 Inverse Problems\
        \else\ifnum\thejnl=6 Class. Quantum Grav.\
        \else\ifnum\thejnl=7 Network\
        \else\ifnum\thejnl=8 Nonlinearity\
        \else\ifnum\thejnl=9 Quantum Opt.\
        \else\ifnum\thejnl=10 Waves in Random Media\
        \else\ifnum\thejnl=11 Pure Appl. Opt.\
        \else\ifnum\thejnl=12 Phys. Med. Biol.\ %
        \else\ifnum\thejnl=13 Modelling Simul.\ Mater.\ Sci.\ Eng.\
        \else\ifnum\thejnl=14 Plasma Phys. Control. Fusion\
        \else\ifnum\thejnl=15 Physiol. Meas.\
        \else\ifnum\thejnl=16 Sov.\ Lightwave Commun.\
        \else\ifnum\thejnl=17 High Perform.\ Polym.\
        \else\ifnum\thejnl=18 J.\ Hard Mater.\
        \else\ifnum\thejnl=19 J.\ Phys.\ D: Appl.\ Phys.\
        \else\ifnum\thejnl=20 Supercond.\ Sci.\ Technol.\
        \else\ifnum\thejnl=21 Semicond.\ Sci.\ Technol.\
        \else\ifnum\thejnl=22 Nanotechnology\
        \else\ifnum\thejnl=23 Meas.\ Sci.\ Technol.\
        \else\ifnum\thejnl=24 Plasma Source Sci.\ Technol.\
        \else\ifnum\thejnl=25 Smart Mater.\ Struct.\
        \else\ifnum\thejnl=26 J.\ Micromech.\ Microeng.\
        \else\ifnum\thejnl=27 Distrib.\ Syst.\ Engng\
\else Institute of Physics Publishing
\fi\fi\fi\fi\fi\fi\fi\fi\fi\fi\fi\fi\fi\fi\fi
\fi\fi\fi\fi\fi\fi\fi\fi\fi\fi\fi\fi}

\def\catchline{\hfill}

\def\cpyrtline{\hfill}
\def\maketitle{\vspace*{\baselineskip}\vspace{0\p@ plus1fil}
    \noindent Short title: \@shorttitle\par
    \@submitted
    \vspace*{\baselineskip}
    \noindent\today\par\newpage}
\def\@rticle#1#2{\thispagestyle{myheadings}%
     \vspace*{.5pc}%
    {\parindent=\mathindent \bf #1\par}%
     \vspace*{1.5pc}%
    {\exhyphenpenalty=10000\hyphenpenalty=10000
     \Large\raggedright\noindent
     \bf#2\par}\def\@shorttitle{#1}\futurelet\next\sh@rttitle}%
\def\title#1{\def\@shorttitle{#1}%
    \thispagestyle{myheadings}%
    \vspace*{3pc}{\exhyphenpenalty=10000\hyphenpenalty=10000
    \Large\raggedright\noindent
    \bf#1\par}\futurelet\next\sh@rttitle}

\def\article#1#2{\@rticle{#1}{#2}}
\def\review#1{\@rticle{REVIEW \ifpmb\else ARTICLE\fi}{#1}}
\def\topical#1{\@rticle{TOPICAL REVIEW}{#1}}
\def\ireview#1{\@rticle{INTRODUCTORY REVIEW}{#1}}
\def\comment#1{\@rticle{COMMENT}{#1}}
\def\note#1{\@rticle{NOTE}{#1}}
\def\prelim#1{\@rticle{PRELIMINARY COMMUNICATION}{#1}}
\def\letter#1{\@rticle{LETTER TO THE EDITOR}{#1}}
\def\sh@rttitle{\ifx\next[\let\next=\sh@rt
                \else\let\next=\f@ll\fi\next}
\def\sh@rt[#1]{\gdef\@shorttitle{#1}}
\def\f@ll{}
\renewcommand{\author}[1]{\vspace*{1.5pc}%
   \begin{indented}%
   \item[]\normalsize\ifnum\thejnl=8\bf\else\rm\fi\raggedright#1
   \end{indented}%
   \smallskip}
\def\abstract{\vspace{16pt plus3pt minus3pt}
   \begin{indented}
   \item[]{\bf \abstractname.}\quad\rm\ignorespaces}%
\def\endabstract{\end{indented}\vspace{18\p@ plus18\p@}}
\def\submitted{\def\@submitted{\vspace{\baselineskip}%
     \noindent Submitted to: \journal\par}}
\def\@submitted{}
\def\nosections{\vspace{30\p@ plus12\p@ minus12\p@}
    \noindent\ignorespaces}
\def\ack{\ifletter\bigskip\noindent\ignorespaces\else
    \section*{Acknowledgments}\fi}

\newif\ifnumbysec
\def\theequation{\ifnumbysec
      \arabic{section}.\arabic{equation}\else
      \arabic{equation}\fi}
\def\eqnobysec{\numbysectrue\@addtoreset{equation}{section}}
\newenvironment{ceqno}{\begin{ceqno}}{\end{ceqno}}
\def\ceqno{\begin{equation}\begin{array}{@{}*{4}{l}}\dsty}
\def\endceqno{\end{array}\end{equation}}
\def\eqalign#1{\null\vcenter{\def\\{\cr}\openup\jot\m@th
  \ialign{\strut$\displaystyle{##}$\hfil&$\displaystyle{{}##}$\hfil
      \crcr#1\crcr}}\,}
\def\eqalignno#1{\displ@y \tabskip\z@skip
  \halign to\displaywidth{\hspace{5pc}$\@lign\displaystyle{##}$%
    \tabskip\z@skip
    &$\@lign\displaystyle{{}##}$\hfill\tabskip\@centering
    &\llap{$\@lign\hbox{\rm##}$}\tabskip\z@skip\crcr
    #1\crcr}}
\def\numparts{\addtocounter{equation}{1}%
     \setcounter{eqnval}{\value{equation}}%
     \setcounter{equation}{0}%
     \def\theequation{\ifnumbysec
     \arabic{section}.\arabic{eqnval}{\it\alph{equation}}%
     \else\arabic{eqnval}{\it\alph{equation}}\fi}}

\def\endnumparts{\def\theequation{\ifnumbysec
     \arabic{section}.\arabic{equation}\else
     \arabic{equation}\fi}%
     \setcounter{equation}{\value{eqnval}}}
\def\cases#1{%
     \left\{\,\vcenter{\def\\{\cr}\normalbaselines\openup1\jot\m@th%
     \ialign{\strut$\displaystyle{##}\hfil$&\tqs
     \rm##\hfil\crcr#1\crcr}}\right.}%
\def\tabular{\def\@halignto{}\@tabular}
\newcommand{\Table}[1]{\def\t@blecap{\caption{#1}}%
   \setcounter{tabtype}{1}\futurelet\next\t@bplace}
\newcommand{\widetable}[1]{\def\t@blecap{\caption{#1}}%
   \setcounter{tabtype}{2}\futurelet\next\t@bplace}
\newcommand{\fulltable}[1]{\def\t@blecap{\caption{#1}}%
   \setcounter{tabtype}{3}\futurelet\next\t@bplace}%
\def\t@bplace{\ifx\next[\let\next=\@tabpl
                 \else\let\next=\@tabnopl\fi\next}
\def\@tabpl[#1]{\begin{table}[#1]\@t@bsize}
\def\@tabnopl{\begin{table}\@t@bsize}
\def\@t@bsize{\ifnum\thetabtype=3\begin{varindent}{0pt}%
   \else\begin{varindent}{\mathindent}\fi
   \t@blecap\lineup\item[]
   \ifnum\thetabtype=1
        \begin{tabular}{@{}l*{15}{l}}
   \else\ifnum\thetabtype=2
        \begin{tabular*}{\indentedwidth}{@{}l*{15}{@{\extracolsep{0pt
plus12pt}}l}}
   \else\begin{tabular*}{\textwidth}{@{}l*{15}{@{\extracolsep{0pt plus12pt}}l}}
   \fi\fi}
\def\endtab{\ifnum\thetabtype=1\end{tabular}
   \else\end{tabular*}\fi\end{varindent}\end{table}}

\def\lineup{\def\0{\hbox{\phantom{\footnotesize\rm 0}}}%
    \def\m{\hbox{$\phantom{-}$}}%
    \def\-{\llap{$-$}}}
\long\def\@makecaption#1#2{\vskip 10\p@
 \ifnum\thefigtype=2\begin{varindent}{\@figindent}
 \item[]{\bf #1.} #2
 \end{varindent}\else
 \ifnum\thefigtype=3
 \footnotesize\rm{\bf #1.} #2\else
 \begin{indented}
 \item[]{\bf #1.} #2
 \end{indented}\fi\fi}
\newcommand{\Figure}[1]{\setcounter{figtype}{1}%
    \def\figspace{}\def\figcap{\caption{#1}}%
    \futurelet\next\@figplace}
\def\@figplace{\ifx\next[\let\next=\@figpl
                 \else\let\next=\@fignopl\fi\next}
\def\@figpl[#1]{\begin{figure}[#1]
   \figspace
   \figcap
   \end{figure}}
\def\@fignopl{\begin{figure}
   \figspace
   \figcap
   \end{figure}}

\newcommand{\sidecap}[3]{\setcounter{figtype}{2}%
    \setlength{\figdepth}{#1}\def\@figindent{#2}%
    \def\sidedc@p{\caption{#3}}%
    \futurelet\next\@sidecapplace}
\def\@sidecapplace{\ifx\next[\let\next=\@sidecappl
                 \else\let\next=\@sidecapnopl\fi\next}
\def\@sidecappl[#1]{\begin{figure}[#1]
    \vbox to\figdepth{\vfill
    \sidedc@p}%
    \setcounter{figtype}{1}\end{figure}}
\def\@sidecapnopl{\begin{figure}
    \vbox to\figdepth{\vfill
    \sidedc@p}%
    \setcounter{figtype}{1}\end{figure}}
\newcommand{\side}[3]{\setcounter{figtype}{3}%
    \setlength{\figdepth}{#1}\setlength{\figwidth}{15pc}
    \setlength{\secfigwidth}{15pc}
    \def\firstc@p{\caption{#2}}\def\secondc@p{\caption{#3}}
    \futurelet\next\@sideplace}
\def\@sideplace{\ifx\next[\let\next=\@sidepl
                 \else\let\next=\@sidenopl\fi\next}
\def\@sidepl[#1]{\begin{figure}[#1]
    \vspace*{1.5pc}\vspace*{\figdepth}
    \parbox[t]{\figwidth}{\firstc@p}\hspace*{1pc}%
    \parbox[t]{\secfigwidth}{\secondc@p}
    \setcounter{figtype}{1}\end{figure}}
\def\@sidenopl{\begin{figure}
    \vspace*{1.5pc}\vspace*{\figdepth}
    \parbox[t]{\figwidth}{\firstc@p}\hspace*{1pc}%
    \parbox[t]{\secfigwidth}{\secondc@p}
    \setcounter{figtype}{1}\end{figure}}
\newcommand{\varside}[4]{\setcounter{figtype}{3}%
    \setlength{\figdepth}{#1}\setlength{\figwidth}{#2}%
    \setlength{\secfigwidth}{30pc}
    \addtolength{\secfigwidth}{-\figwidth}
    \def\firstc@p{\caption{#3}}\def\secondc@p{\caption{#4}}
    \futurelet\next\@sideplace}

\newcounter{figure}
\def\thefigure{\@arabic\c@figure}
\def\fps@figure{htbp}
\def\ftype@figure{1}
\def\ext@figure{lof}
\def\fnum@figure{\figurename~\thefigure}
\def\figure{\@float{figure}}
\let\endfigure\end@float
\@namedef{figure*}{\@dblfloat{figure}}
\@namedef{endfigure*}{\end@dblfloat}
\newcounter{table}
\def\thetable{\@arabic\c@table}
\def\fps@table{htbp}
\def\ftype@table{2}
\def\ext@table{lot}
\def\fnum@table{\tablename~\thetable}
\def\table{\@float{table}}
\let\endtable\end@float
\@namedef{table*}{\@dblfloat{table}}
\@namedef{endtable*}{\end@dblfloat}

\def\thebibliography#1{\list
 {\hfil[\arabic{enumi}]}{\topsep=0\p@\parsep=0\p@
 \partopsep=0\p@\itemsep=0\p@
 \labelsep=5\p@\itemindent=-10\p@
 \settowidth\labelwidth{\footnotesize[#1]}%
 \leftmargin\labelwidth
 \advance\leftmargin\labelsep
 \advance\leftmargin -\itemindent
 \usecounter{enumi}}\footnotesize
 \def\newblock{\ }
 \sloppy\clubpenalty4000\widowpenalty4000
 \sfcode`\.=1000\relax}

\def\numrefs#1{\begin{thebibliography}{#1}}
\def\endnumrefs{\end{thebibliography}}

\def\thereferences{\list{}{\topsep=0\p@\parsep=0\p@
 \partopsep=0\p@\itemsep=0\p@\labelsep=0\p@\itemindent=-18\p@
\labelwidth=0\p@\leftmargin=18\p@
}\footnotesize\rm
\def\newblock{\ }
\sloppy\clubpenalty4000\widowpenalty4000
\sfcode`\.=1000\relax
}
\newenvironment{harvard}{\list{}{\topsep=0\p@\parsep=0\p@
\partopsep=0\p@\itemsep=0\p@\labelsep=0\p@\itemindent=-18\p@
\labelwidth=0\p@\leftmargin=18\p@
}\footnotesize\rm
\def\newblock{\ }
\sloppy\clubpenalty4000\widowpenalty4000
\sfcode`\.=1000\relax}{\endlist}
\def\refs{\begin{harvard}}
\def\endrefs{\end{harvard}}
\newenvironment{indented}{\begin{indented}}{\end{indented}}
\newenvironment{varindent}[1]{\begin{varindent}{#1}}{\end{varindent}}
\def\indented{\list{}{\itemsep=0\p@\labelsep=0\p@\itemindent=0\p@
   \labelwidth=0\p@\leftmargin=\mathindent\topsep=0\p@\partopsep=0\p@
   \parsep=0\p@\listparindent=15\p@}\footnotesize\rm}

\def\varindent#1{\setlength{\varind}{#1}%
   \list{}{\itemsep=0\p@\labelsep=0\p@\itemindent=0\p@
   \labelwidth=0\p@\leftmargin=\varind\topsep=0\p@\partopsep=0\p@
   \parsep=0\p@\listparindent=15\p@}\footnotesize\rm}

\def\tabnotes{\ifnum\thetabtype=1\end{tabular}\else\end{tabular*}\fi}
\def\endtabnotes{\end{varindent}\end{table}}

\newif\if@restonecol
\def\theindex{\@restonecoltrue\if@twocolumn\@restonecolfalse\fi
\columnseprule \z@
\columnsep 35\p@\twocolumn[\section*{\indexname}]%
    \@mkboth{{\indexname}}{{\indexname}}%
    \parindent\z@
    \parskip\z@ plus.3\p@\relax\let\item\@idxitem}
\def\@idxitem{\par\hangindent 30\p@}
\def\subitem{\par\hangindent 30\p@ \hspace*{10\p@}}
\def\subsubitem{\par\hangindent 30\p@ \hspace*{20\p@}}
\def\endtheindex{\if@restonecol\onecolumn\else\clearpage\fi}
\def\indexspace{\par \vskip 10\p@ plus 5\p@ minus 3\p@\relax}

\mark{{}{}}

\def\ps@headings{\let\@mkboth\markboth
 \def\@oddfoot{}%
 \def\@evenfoot{}%
 \def\@evenhead{\makebox[\mathindent][l]{\normalsize\rm \thepage}%
  \normalsize\it\rightmark\hfill}%
 \def\@oddhead{\makebox[\mathindent][r]{\hfill}{\normalsize\it\leftmark}\hfill
  \normalsize\rm\thepage}%
}%

\def\ps@myheadings{\let\@mkboth\markboth
 \def\@oddhead{\catchline}%
 \def\@oddfoot{\cpyrtline}%
 \def\@evenhead{}%
 \def\@evenfoot{}%
}

\def\today{\ifcase\month\or
 January\or February\or March\or April\or May\or June\or
 July\or August\or September\or October\or November\or December\fi
 \space\number\day, \number\year}

\def\@begintheorem#1#2{\rm \trivlist \item[\hskip \labelsep{\it #1\ #2.}]}
\def\@opargbegintheorem#1#2#3{\rm \trivlist
      \item[\hskip \labelsep{\it #1\ #2\ (#3).}]}

\def\p@LaTeX{{L\kern-.3em\lower.1em\hbox{$^{\rm A}$}\kern-.15em%
    T\kern-.1667em\lower.7ex\hbox{E}\kern-.125emX}}
\newcommand{\mathrm}{\rm}
\newcommand{\text}[1]{\mbox{#1}}

\newcommand{\nohyphens}{\hyphenpenalty=10000\exhyphenpenalty=10000}

\renewcommand{\qquad}{\hspace*{25pt}}
\newcommand{\tqs}{\hspace*{25pt}}
\newcommand{\fl}{\hspace*{-\mathindent}}

\def\pt(#1){({\it #1\/})}
\newcommand{\dsty}{\displaystyle}
\newcommand{\tsty}{\textstyle}

\def\;{\protect\psemicolon}
\def\psemicolon{\relax\ifmmode\mskip\thickmuskip\else\kern .3333em\fi}

\newcommand{\opencirc}{\raisebox{2\p@}{\;\circle{5}}}

\newcommand{\fullcirc}{\raisebox{-2\p@}{\Large$\bullet$}}

\newcommand{\boldarrayrulewidth}{1\p@}

\def\bhline{\noalign{\ifnum0=`}\fi\hrule \@height
\boldarrayrulewidth \futurelet \@tempa\@xhline}

\def\@xhline{\ifx\@tempa\hline\vskip \doublerulesep\fi
      \ifnum0=`{\fi}}

\newcommand{\ms}{\noalign{\vspace{3\p@ plus2\p@ minus1\p@}}}
\newcommand{\bs}{\noalign{\vspace{6\p@ plus2\p@ minus2\p@}}}
\newcommand{\ns}{\noalign{\vspace{-3\p@ plus-1\p@ minus-1\p@}}}
\newcommand{\es}{\noalign{\vspace{6\p@ plus2\p@ minus2\p@}}\displaystyle}
\newcommand{\JPA}{{\em J. Phys. A: Math. Gen.} }
\newcommand{\JPC}{{\em J. Phys. C: Solid State Phys.} }   
\newcommand{\JPCM}{{\em J. Phys.: Condens. Matter\/} }    

\newcommand{\JMP}{{\em J. Math. Phys.} }

\newcommand{\NP}{{\em Nucl. Phys.} }
\newcommand{\PL}{{\em Phys. Lett.} }
\newcommand{\PR}{{\em Phys. Rev.} }
\newcommand{\PRL}{{\em Phys. Rev. Lett.} }

\ps@headings  \onecolumn
\makeatother
\begin{document}
\jl{1}
\def\cs{a}
\def\ca{b}
\def\tl{$T_{N}(q)$}
\def\H{H^{(N)}}
\begin{center}
   {\bf \huge Temperley--Lieb Words \\
     as Valence-Bond Ground States}
   \\[25MM]
Peter F. Arndt\dag , Thomas Heinzel\dag\  and C.M. Yung\ddag
\\[7mm]
{\it \dag Physikalisches Institut \\ Universit\"at Bonn, Nussallee 12,
53115 Bonn, Germany \\[5mm]
  \ddag Department of Mathematics \\ Australian National
 University,Canberra, ACT 0200, Australia}
\\[2.2cm]
{\bf Abstract}
\end{center}
\renewcommand{\thefootnote}{\arabic{footnote}}
\addtocounter{footnote}{-1}
\vspace*{2mm}
%
Based on the Temperley--Lieb algebra we define
a class of one-dimensional Hamiltonians with nearest and
next-nearest neighbour interactions.
Using the regular representation we give ground states of this model
as words of the algebra.
Two point
correlation functions can be computed
employing the Temperley--Lieb relations.
Choosing a spin-$\frac12$ representation of the algebra
we obtain a generalization of the
($q$-deformed) Majumdar--Ghosh model.
The ground states become valence-bond states.
\vspace*{4.0cm}
\begin{flushleft}
BONN HE-94-26 \\
cond-mat/9411085
\\
November 1994
\end{flushleft}
\thispagestyle{empty}
\mbox{}
\newpage
\setcounter{page}{1}
\noindent In this paper we present a class of
one-dimensional Hamiltonians
$H=H(\cs ,\ca ,q)$
with nearest and next-nearest
neighbour interaction.
The Hamiltonian is given in terms of elements
of a Temperley--Lieb algebra \cite{TemLieb} and
has the structure of the Majumdar--Ghosh model
\cite{MaGh,MaGh2,Broe,Aff,BatYung}.
With specific representations of this algebra
one obtains various quantum spin chains.
The functions $\cs=\cs(q)$ and $\ca=\ca(q)$
determine the next-nearest
neighbour interaction and can be chosen
such that the ground state can be given explicitly.
In a graphical form of the regular representation of
the Temperley--Lieb algebra \cite{MarB,Hin}
these ground states have a particular simple form.
They are related to valence-bond spin states.
We also
calculate correlation functions using the graphical
representation.

A Temperley--Lieb algebra \tl\
is defined by the following
relations on the generators $e_i$, $i=1,2,\ldots,N-1$ :
\numparts
\begin{eqnarray}
&e_i\; e_i=x\; e_i=(q+q^{-1})\; e_i \label{tl1}\\
&e_i\; e_{i \pm 1}\; e_i=e_i \label{tl2}\\
&e_i\; e_j=e_j\; e_i \label{tl3} \hspace{2cm} (j \neq i\pm 1)
\end{eqnarray}
\endnumparts
Here we consider the case where $q$ is real.
These algebras appear as centralizer algebras of
the quantum group $U_qSU(2)$ \cite{MarB,PasSal}.
Therefore $U_qSU(2)$ invariant models
naturally show an underlying Temperley--Lieb structure.
However, our results can be used for models
having other quantum group symmetries as well
\cite{BatKun,BatMezNepRit,Affe,BarBat}.

\vspace{2ex}
We define the Hamilton operator
as an abstract element of the Temperley--Lieb algebra.
We use the two-point correlation operators defined
in \cite{Hin}.
These operators preserve the quantum group symmetry
of the respective representations of
the Temperley--Lieb algebra.
Two types of two-point operators can be defined by the
recursive relations
\begin{eqnarray}
g_{l,l+1}&=g^{\pm}_{l,l+1}=e_l-(q+q^{-1})^{-1}
                 &1\leq l\leq N-1 \nonumber\\
g^{\pm}_{l,m}&=-q^{\pm 1}g^{\pm}_{l,n}g^{\pm}_{n,m}
                  -q^{\mp 1}g^{\pm}_{n,m}g^{\pm}_{l,n}
                  \qquad \qquad
                 &1\leq l<n<m\leq N \label{Corr}\\
g^{\pm}_{m,l}&=q^{\mp 4} g^{\pm}_{l,m}
                 &1\leq l<m\leq N.\nonumber
\end{eqnarray}
In the following we use only $g^{\pm}_{l,m}$ with $l<m$.
Note that the definition is in terms of generators $e_l$ of
{\tl}, independent of their realization.

We employ both $g^{+}_{l,m}$ and $g^{-}_{l,m}$ to define
the Hamilton operator
\begin{equation}
\label{H}
H^{(N)}(\cs,\ca,q)= \frac{2-x^2}{2x(x^2-1)} \sum_{i=1}^{N-2}
h_{i,i+1,i+2}+2\cs(q) (N-2){(2-x^2)^2}
\end{equation}
with
\begin{eqnarray}
\label{hi}
h_{i,i+1,i+2} =& 2g_{i,i+1} +2((x^2-2)\cs(q) -1)g_{i+1,i+2}
\nonumber \\
       &+\cs(q) (g_{i,i+2}^+ + g_{i,i+2}^-)
       +\ca(q) (g_{i,i+2}^+ - g_{i,i+2}^-).
\end{eqnarray}
The operator $h_{i,i+1,i+2}$ acts on sites $i$, $i+1$, $i+2$
and involves the generators $e_i$ and $e_{i+1}$.
The value of $x$ is given by the Temperley--Lieb relation
(\ref{tl1}).
The arbitrary functions $\cs=\cs(q)$ and $\ca=\ca(q)$
are weighting factors of the symmetric and antisymmetric
contribution of $g_{i,i+2}^+$ and $g_{i,i+2}^-$ to the Hamiltonian.
For $q=1$ the Hamiltonian is independent of $\ca$, since
$g^+_{l,m}=g^-_{l,m}$.
It is useful to notice how the Hamiltonian changes, if we
replace $q$ by $1/q$.
We have the identity
\begin{equation}
\label{symm}
H^{(N)}(\cs,\ca,q)=H^{(N)}(\cs,-\ca,1/q)  \;.
\end{equation}
The coefficients in (\ref{hi}) are chosen such that
simple ground states can be found.
To this end we use the regular representation of
\tl\ on boundary diagrams \cite{MarB}.
We find the ground states to be specific
Temperley--Lieb words.
A word in this context means a straight product
of generators and is realized by a single boundary
diagram.

A boundary diagram is given by two rows of $N$ upper and $N$ lower
points with $N$ non-intersecting lines connecting the points
such that any point is connected to a single other point.
A generator is realized as
%
%
\setlength{\unitlength}{6pt}
\thicklines
\def\sgr{\scriptscriptstyle}
\def\punkte{\begin{picture}(0,3)\multiput(0,0)(1,0){3}{\circle*{0.1}}
            \end{picture}}
\def\linie{\begin{picture}(0,6)\put(0,0){\line(0,1){6}}\end{picture}}
\def\bogenu{\begin{picture}(2,2)\put(0,0){\oval(2,2)[t]}\end{picture}}
\def\bogeno{\begin{picture}(1,2)\put(0,0){\oval(2,2)[b]}\end{picture}}
\def\ei{\begin{picture}(18,6)(0,2.7)
        \put(0,0){\linie} \put(2,0){\linie}
        \put(3,3){\punkte} \put(6,0){\linie}
        \put(9,0){\bogenu} \put(9,6){\bogeno}
        \put(12,0){\linie} \put(13,3){\punkte}
        \put(16,0){\linie} \put(18,0){\linie}
        \put(0,-0.4){\makebox(0,0)[t]{$\sgr 1$}}
        \put(2,-0.4){\makebox(0,0)[t]{$\sgr 2$}}
        \put(8,-0.4){\makebox(0,0)[t]{$\sgr i$}}
        \put(10,-0.4){\makebox(0,0)[t]{$\sgr i+1$}}
        \put(16,-0.4){\makebox(0,0)[t]{$\sgr N-1$}}
        \put(18,-0.4){\makebox(0,0)[t]{$\sgr N$}}
        \end{picture}}
\begin{equation}
\label{bdgen}
e_i=\; \ei \; \;.
\end{equation}
\vspace{0.6cm}
The composition of two words is defined as stacking the corresponding
diagrams on top of each other and identifying
the lower points of the first with the upper points of the
second diagram.
Any closed line appearing in this process is discarded from
the diagram and replaced by a factor $x=q+q^{-1}$.
This reflects the relation (\ref{tl1}).
The other Temperley--Lieb relations (\ref{tl2}) and (\ref{tl3})
can easily be verified by drawing the corresponding diagrams.
By successive composition of generators (\ref{bdgen})
all distinct words of {\tl} are realized as different
boundary diagrams.
Thus for calculations in the regular representation
of {\tl} we can take all
possible boundary diagrams as a basis of the representation space.

In the following we give eigenvectors and eigenvalues of $\H $.
First we take $N$ even.
Employing boundary diagrams it is straightforward to show
that (\ref{H}) has the eigenvector
%
%
\def\vone{\begin{picture}(14,6)(0,2.7)
        \put(1,0){\bogenu}
        \put(5,0){\bogenu}
        \put(13,0){\bogenu}
        \put(1,6){\bogeno}
        \put(5,6){\bogeno}
        \put(13,6){\bogeno}
        \put(7.5,5.5){\punkte}
        \put(7.5,0.5){\punkte}
        \end{picture}}
\begin{equation}
\label{v1}
v_1=e_1e_3\ldots e_{N-1}\; =\;  \vone \; \; .
\end{equation}
\vspace{0.6cm}
The additive constant in (\ref{H}) is chosen such that
its eigenvalue is zero.

If we impose a condition
on the functions $\cs(q)$ and $\ca(q)$:
\begin{equation}
\label{cond}
\cs=(2-x(q-q^{-1})\ca)/(x^2-1),
\end{equation}
we find two further eigenwords with eigenvalue zero:
%
%
\def\vtwo{\begin{picture}(16,6)(0,2.7)
        \put(1,0){\bogenu}
        \put(5,0){\bogenu}
        \put(15,0){\bogenu}
        \put(3,6){\bogeno}
        \put(7,6){\bogeno}
        \put(13,6){\bogeno}
        \put(8,6){\oval(16,5)[b]}
        \put(9,0.5){\punkte}
        \put(9,5.5){\punkte}
        \end{picture}}
\def\vthree{\begin{picture}(16,6)(0,2.7)
        \put(0,0){\linie}
        \put(3,0){\bogenu}
        \put(7,0){\bogenu}
        \put(3,6){\bogeno}
        \put(7,6){\bogeno}
        \put(9,3){\punkte}
        \put(13,0){\bogenu}
        \put(13,6){\bogeno}
        \put(16,0){\linie}
        \end{picture}}
\begin{equation}
\label{v2}
v_2=(e_2e_4\ldots e_{N-2})\; v_1\;=\; \vtwo
\end{equation}
\vspace{0.3cm}
and
\begin{equation}
\label{v3}
v_3=e_2e_4\ldots e_{N-2}\;=\; \vthree \; \; .
\end{equation}
\vspace{0.6cm}

For an odd number of sites $N$ and the condition
(\ref{cond}) we have two eigenwords that
are given by the diagram (\ref{v3})
without the left or right vertical line.
We denote these words as $v_4=e_1e_3\ldots e_{N-2}$
respectively $v_5=e_2e_4\ldots e_{N-1}$.
Their eigenvalues can be computed as
$\mp{\ca}/2\,(q^2+q^{-2})(q-q^{-1}){(q+q^{-1})}^{-1}$ for $v_4$
respectively $v_5$.

For the following we impose condition (\ref{cond}).
We investigate whether we can choose the function
$\ca(q)$ such that the eigenstates $v_i$ given above become ground states
of the Hamiltonian $\H$.
We can show that $\H$ has a ground state $v_i$, if
the function $\ca(q)$ is bounded by
\begin{equation}
\label{condgs}
\ca\;\leq (\geq) \;
\frac{q^3+2q+2q^{-1}+q^{-3}}{q^5+2q^3-2q^{-3}-q^{-5}}
\qquad \mbox{for }|q|>1 \;(|q|<1)
\end{equation}
The relation between the cases $|q|>1$ and $|q|<1$
reflects the symmetry (\ref{symm}).

For a proof first consider the case $N$ even.
One can think of $\H$ as the sum of $N/2-1$
Hamilton operators $H^{(4)}$
each involving only three generators.
This can be seen by grouping together $h_{i,i+1,i+2}$ and
$h_{i+1,i+2,i+3}$ ($i$ odd) in the definition (\ref{H}).
The operator $H^{(4)}$ can be diagonalized (see below).
Hence the lowest eigenvalue of $\H$ is bounded from below by
$N/2-1$ times the lowest eigenvalue of $H^{(4)}$.
We take the $14$ possible boundary diagrams with $N=4$
as a basis to write $H^{(4)}$
in the regular representation.
Diagonalization of this $14\times 14$ matrix gives the eigenvalues
\begin{eqnarray}
\fl\lambda_1&=&0 \nonumber\\
\fl\lambda_2&=&2-(q-q^{-1})\ca(q) \nonumber\\
\fl\lambda_3&=&1-(q-q^{-1}){(q+q^{-1})}^{-1}\ca(q) \nonumber\\
\fl\lambda_4&=&1-(q^3-q^{-3}){(q+q^{-1})}^{-1}\ca(q) \nonumber\\
\fl\lambda_5&=&1-(q^5+2q^{3}-2q^{-3}-q^{-5})
  {(q^{3}+2q+2q^{-1}+q^{-3})}^{-1}\ca(q)
  \nonumber\\
\fl\lambda_{6/7}&=&1-\frac12{(q^{3}+2q+2q^{-1}+q^{-3})}^{-1}\;
  \bigg[ (q^5+2q^{3}+q-q^{-1}-2q^{-3}-q^{-5})\ca(q) \nonumber\\
\fl &&\pm \Big[ (q^{10}+6q^{6}-2q^{4}+q^{2}-12+
  q^{-2}-2q^{-4}+6q^{-6}+q^{-10}){\ca}^2(q) \nonumber\\
\fl &&-8(q^{4}+q^{2}-q^{-2}-q^{-4})\ca(q) +4(q+q^{-1})^2 \Big]
  ^{\frac12}\bigg] \;\;\;.\nonumber
\end{eqnarray}
The degeneracy of the first eigenvalue is seven and of the second
two.
The others are non-degenerate.
{}From this result we can conclude that, if we have equation
(\ref{condgs}), all eigenvalues of $H^{(4)}$
are greater or equal to zero.
Hence for functions $\ca(q)$, that fulfill conditions (\ref{cond})
and (\ref{condgs}),
$v_1$,$v_2$ and $v_3$ are ground states of the
Hamiltonian $\H$ because their eigenvalue is zero.

For $N$ odd we can analogously view $\H$ as the sum of $(N-3)/2$
operators $H^{(4)}$ and one operator $H^{(3)}$.
Making use of the $5$ different boundary diagrams for $N=3$
we find the eigenvalues of $H^{(3)}$:
\begin{eqnarray}
\mu_1=1-\frac{\ca(q)}{2}(q^2-q^{-2}) \nonumber\\
\mu_{2/3}=\pm\frac{\ca(q)}{2}(q^2+q^{-2})(q-q^{-1}){(q+q^{-1})}^{-1}.
     \nonumber
\end{eqnarray}
The eigenvalues $\mu_2$ and $\mu_3$ are twofold degenerate,
$\mu_1$ is non-degenerate.
If we impose condition (\ref{condgs}),
$\mu_1$ is always positive.
Note that for (\ref{cond}) the value
$\mu_2$ ($\mu_3$) is also an eigenvalue of $\H$
with eigenvector $v_4$ ($v_5$).
Thus imposing both conditions (\ref{cond}) and (\ref{condgs})
a ground state of $\H$ is given by $v_4$ or $v_5$ depending on the
values of $q$ and $\ca$.
Explicitely, for $|q|\geq 1$ and $\ca\leq 0$ ($\ca\geq 0$) we
have $v_4$ ($v_5$)
as ground state of $\H$.
For $|q|\leq1$ and $\ca\leq 0$ ($\ca\geq 0$) we find
the ground state $v_5$ ($v_4$).

If we drop condition (\ref{cond}), $v_1$ remains eigenstate of
$\H$ for $N$ even.
For this case we want to remark,
that one can choose two
functions $\cs(q)$ and $\ca(q)$ within certain bounds
to make $v_1$ a ground state of the Hamiltonian.

Next we calculate correlation functions for the ground
states $v_1$ and $v_2$.
Again the computation is completely general in terms of boundary
diagrams.

For a graphical calculation of the correlation functions
we have to restrict our attention to the words of the
left-sided ideal that is generated by $v_1=e_1e_3\ldots e_{N-1}$
($N$ even).
The diagrams with lower part as in $v_1$ and arbitrary
upper part
constitute this ideal $\cal S$.
Thus the ground states $v_1$ and $v_2$ belong to $\cal S$.

Defining a transposed diagram as the up-side-down reflected
diagram, the scalar product $\langle s_1|s_2 \rangle$
of two elements $s_1$ and $s_2$ of $\cal S$
can be defined as follows \cite{Hin}.
If one removes any loop from the diagram representing
${s_1}^T\; s_2$, the resulting
diagram is always $v_1$ with a factor depending on $s_1$ and $s_2$.
The scalar product is given by this factor:
\begin{equation}
\label{picsp}
{s_1}^T\; s_2\; = \; \langle s_1|s_2 \rangle \; \vone \;.
\end{equation}
\vspace{0.6cm}
For example it is easy to see that
\begin{equation}
\label{norm}
\langle v_1|v_1 \rangle=\langle v_2|v_2 \rangle=x^{N/2}  \;.
\end{equation}
Thus we can calculate
the expectation value of the
correlation operators $g^{\pm}_{l,n}$ for a state $v$ in $\cal S$.
We evaluate the diagrams
\begin{equation}
\label{gpic}
v^T\; g^{\pm}_{l,m}\; v= \langle v|g_{l,m}|v \rangle \; \vone \; \;
\end{equation}
\vspace{0.6cm}
using the recurrence relation
(\ref{Corr}) expressed in terms of diagrams.
This way by induction on k,
we can show that the correlation
functions for the states $v_1$ and $v_2$ for all $N$ are given by
\begin{equation}
\langle v_i |g^{\pm}_{l,l+k}| v_i \rangle = \left\{ \begin{array}{ll}
            (x-\frac1x)\,x^{N/2}
                             &\mbox{for $i=1$, $k=1$ and $l$ odd} \\
            \ms (x-\frac1x)\,x^{N/2}
                             &\mbox{for $i=2$, $k=1$ and $l$ even} \\
                             &\mbox{and for $i=2$, $l=1$, $l+k=N$} \\
            \ms 0 &\mbox{otherwise.}\\
            \end{array} \right.
\end{equation}
Thus only sites connected by an upper line in $v_{1/2}$
have a non-zero correlation.

Note that we find trivial short range correlations,
although the
operators $g^{\pm}_{l,l+k}$ have a non-local structure
(\ref{Corr}).
%
%
%
%
The ground states do not have any
long range order.
This can be expected from the interpretation of the words $v_1$
and $v_2$ in terms of spins that we give in the following paragraph
for the representation (\ref{rep}) with spin $\frac12$.

\vspace{2ex}
So far we have used the regular representation of the
Temperley--Lieb algebra producing results
independent of special representations.
We now take the $U_qSU(2)$ spin-$\frac12$ representation of \tl,
that is defined by \cite{Hin,PasSal}
\begin{equation}
\label{rep}
\fl e_l = -\frac12\Bigl(
       \sigma_l^x\sigma_{l+1}^x+\sigma_l^y\sigma_{l+1}^y+
       \frac{q+q^{-1}}{2}(\sigma_l^z\sigma_{l+1}^z-1)+
       \frac{q-q^{-1}}{2}(\sigma^z_l-\sigma^z_{l+1}) \Bigr) \;
\end{equation}
where $\sigma^x$, $\sigma^y$ and $\sigma^z$ are Pauli matrices.
For this representation  the $g^{\pm}_{l,m}$ (\ref{Corr})
are $U_qSU(2)$ invariant
generalizations of the $SU(2)$ invariant scalar product \cite{Hin}
\begin{equation}
g_{l,m}^{q=1}=-\frac12 \vec{\sigma}_l \cdot \vec{\sigma}_m.
\end{equation}

The resulting Hamiltonian $\H$ is $U_qSU(2)$ invariant.
One would like to know the spin configuration states
that correspond to the boundary diagrams $v_i$, i.e. the
ground states of this quantum chain for conditions
(\ref{cond}) and (\ref{condgs}).
To this end consider the special case
\begin{equation}
\label{condproj}
\ca(q)=0 \mbox{  and  } \cs(q)=2/(x^2-2)\;\;.
\end{equation}
With the chosen representation and this condition
$\H$ becomes the Hamiltonian of the $q$-deformed
Majumdar--Ghosh model \cite{MaGh,BatYung} given by
\begin{equation}
\label{HMG}
H^{\rm MG}=\sum_{i=1}^{N-2} P_{i,i+1,i+2}^{3/2}\;\;
\end{equation}
where $P_{i,i+1,i+2}^{3/2}$ is the projector onto the
($q$-deformed) quartet of the spins at site $i$, $i+1$ and $i+2$.

The Majumdar--Ghosh model is known to have a
valence-bond ground state \cite{MaGh2,Broe,BatYung}.
Denoting the spin-$\frac12$ representation space of a site
by a dot and a singlet combination of two adjacent spins
by a short line,
they can be given pictorially as
\setlength{\unitlength}{6.0mm}
\thicklines
\def\dma{0.3}
\def\dmb{0.1}
\def\wone{\begin{picture}(10,1)(0,-0.2)
        \put(1,0){\circle*{\dma}}
        \put(2,0){\circle*{\dma}}
        \put(3,0){\circle*{\dma}}
        \put(4,0){\circle*{\dma}}
        \put(5,0){\circle*{\dma}}
        \put(6,0){\circle*{\dma}}
        \put(9,0){\circle*{\dma}}
        \put(10,0){\circle*{\dma}}
        \put(7,0){\circle*{\dmb}}
        \put(7.5,0){\circle*{\dmb}}
        \put(8,0){\circle*{\dmb}}
	\put(1,0){\line(1,0){1}}
	\put(3,0){\line(1,0){1}}
	\put(5,0){\line(1,0){1}}
	\put(9,0){\line(1,0){1}}
        \end{picture}}
\def\wtwo{\begin{picture}(10,1)(0,-0.2)
        \put(1,0){\circle*{\dma}}
        \put(2,0){\circle*{\dma}}
        \put(3,0){\circle*{\dma}}
        \put(4,0){\circle*{\dma}}
        \put(5,0){\circle*{\dma}}
        \put(8,0){\circle*{\dma}}
        \put(9,0){\circle*{\dma}}
        \put(10,0){\circle*{\dma}}
        \put(6,0){\circle*{\dmb}}
        \put(6.5,0){\circle*{\dmb}}
        \put(7,0){\circle*{\dmb}}
	\put(2,0){\line(1,0){1}}
	\put(4,0){\line(1,0){1}}
	\put(8,0){\line(1,0){1}}
        \end{picture}}
\begin{eqnarray}
\label{w1}
w_1=\; \wone \\
\label{w2}
w_2=\; \wtwo \; \; .
\end{eqnarray}
for an even number of sites.
For $N$ odd the ground state has the form of (\ref{w2})
without the dot to the very left or very right.

One can easily verify that $w_1$ and $w_2$ are in fact ground states.
The action of any $P_{i,i+1,i+2}^{3/2}$ on either of these states gives
zero, since two of the three spins of sites $i,i+1$ and $i+2$
are in a singlet configuration.
Thus both states are eigenvectors of $H^{\rm MG}$ with
zero eigenvalue.
Also all possible eigenvalues of $H^{\rm MG}$
are greater or
equal to zero, because it is a sum of projectors.

On the other hand we have the ground states $v_1$,$v_2$ and $v_3$
(respectively $v_4$ or $v_5$ for $N$ odd),
where $v_1$ and $v_2$ belong to the ideal $\cal S$.
For the representation (\ref{rep})
$\cal S$ is known to represent the $U_qSU(2)$ scalar states
\cite{MarB,Hin}.

At this point the correspondence
between the diagrams $v_i$ and spin configurations $w_j$
becomes clear.
A line connecting two upper points in a boundary diagram is
represented by the singlet configuration of the spins at
the corresponding sites.
For this identification one can easily show
that the action of the Temperley--Lieb algebra
on a diagram of $\cal S$ is mirrored by the action of
matrices (\ref{rep}) on the matching spin state.
In this way the two words $v_1$ and $v_2$ are represented by
the two $U_qSU(2)$ ground state singlets
$w_1$ respectively $w_2$ (with the spins of sites $1$ and $N$
in the singlet combination).
The correct normalization
of the spin states can be calculated from (\ref{norm}).
Further analyzing the action of {\tl}
we find that the word $v_3$ corresponds to
a linear combination of the
ground state triplet and the singlet that are both
pictorially described by $w_2$.
Finally for $N$ odd, $v_4$ and $v_5$
correspond to $w_2$ without the left respectively right dot.

\vspace{2ex}

Recently generalizations of the Majumdar-Ghosh model have been
discussed.
A class of $SU(2)$ symmetric antiferromagnetic chains
with valence-bond ground state
can be found in \cite{ShaSut}.
Takano has introduced a generalization of the projection
operators which appear in the formulation of the
Majumdar--Ghosh model \cite{Ta}.
A $q$-deformation of the $SU(2)$ symmetric model
has been given in \cite{BatYung}.
With representation (\ref{rep}) the class of Hamiltonians $\H$
is an extension of the latter $U_qSU(2)$
symmetric Majumdar--Ghosh chain.
Note that
the valence-bond ground states remain unchanged for
a range of functions $\cs(q)$ and $\ca(q)$.
A similar phenomenon was already found for a spin-1 chain
with matrix product ground state \cite{Klue}.
We have not investigated, whether our Hamiltonian is
massless or not \cite{Aff}.

Choosing a different representation we can define further models
with the same property.
It is possible
to represent a Temperley--Lieb algebra on
quantum chains with $n=2s+1$ states per site.
Such a representation of \tl\
is given through the matrix elements \cite{BatMezNepRit}
\begin{eqnarray}
\label{repn}
\langle m_i,m_{i+1} |\; e_i\; | m_i',m_{i+1}' \rangle =
       (-1)^{m_i-m_i'} \; p^{m_i+m_i'} \;\delta_{m_i+m_{i+1},0}
        \;\delta_{m_i'+m_{i+1}',0}
\end{eqnarray}
where $m_i$ is the spin variable at site $i$
with $-s\leq m_i \leq s$.
The value of $p$ can be calculated from
\begin{equation}
q+q^{-1}=x=[n]_p\equiv (p^n-p^{-n})(p-p^{-1})^{-1}\;.
\end{equation}
For $s=1/2$ this reduces to the representation
given in (\ref{rep}).
The resulting Hamiltonians $\H$ are
$U_pSU(n)$ symmetric \cite{BatMezNepRit,Affe}.
In general the matrices (\ref{repn}) realize the projector
onto the $U_pSU(n)$ singlet at two
adjacent sites \cite{BatKun,BatMezNepRit} which is realized
via the branching rule
\begin{equation}
n\times \bar{n}= (n^2-1)+1\;.  \nonumber
\end{equation}
Thus even for $s>1/2$ the ground state is of
valence-bond type.
For example we can expect the ground state of the
$U_pSU(3)$ symmetric model to be tenfold
degenerate (one octet and two singlets).

We have given the correspondence between specific words of
the Temperley--Lieb algebra
and vectors of the spin-$\frac12$ configuration space.
The general relation of words of \tl\ and spin states will
be discussed in a future publication.

\ack
We would like to thank V.~Rittenberg for suggesting
the problem.
Thanks also to M.T.~Batchelor and F.H.L.~E\ss ler for
many helpful discussions.

%

\end{document}